# A second-order orientation-contrast stimulus for population-receptive-field-based retinotopic mapping

Funda Yildirim*[(1)], Joana Carvalho[(1)], Frans W. Cornelissen[(1)]

[(1)] Laboratory of Experimental Ophthalmology, University Medical Center Groningen, University of Groningen, Netherlands

## Abstract

Visual field or retinotopic mapping is one of the most frequently used paradigms in fMRI. It uses activity evoked by position-varying high luminance contrast visual patterns presented throughout the visual field for determining the spatial organization of cortical visual areas. While the advantage of using high luminance contrast is that it tends to drive a wide range of neural populations – thus resulting in high signal-to-noise BOLD responses – this may also be a limitation, especially for approaches that attempt to squeeze more information out of the BOLD response, such as population receptive field (pRF) mapping. In that case, more selective stimulation of a subset of neurons – despite reduced signals – could result in better characterization of pRF properties. Here, we used a second-order stimulus based on local differences in orientation texture – to which we refer as orientation contrast – to perform retinotopic mapping. Participants in our experiment viewed arrays of Gabor patches composed of a foreground (a bar) and a background. These could only be distinguished on the basis of a difference in patch orientation. In our analyses, we compare the pRF properties obtained using this new orientation contrast-based retinotopy (OCR) to those obtained using classic luminance contrast- based retinotopy (LCR). Specifically, in higher order cortical visual areas such as LO, our novel approach resulted in non-trivial reductions in estimated population receptive field size of around 30%. A set of control experiments confirms that the most plausible cause for this reduction is that OCR mainly drives neurons sensitive to orientation contrast. We discuss how OCR – by limiting receptive field scatter and reducing BOLD displacement – may result in more accurate pRF localization as well. Estimation of neuronal properties is crucial for interpreting cortical function. Therefore, we conclude that using our approach, it is possible to selectively target particular neuronal populations, opening the way to use pRF modeling to dissect the response properties of more clearly-defined neuronal populations in different visual areas.

# Introduction

Retinotopic mapping uses the time-varying position of the patterns in the visual field to localize the borders and determining the spatial organization of visual areas. Recent approaches in fMRI analysis and retinotopic mapping have turned to using biologically plausible model-based analyses to reveal more detailed properties of visual areas and neuronal populations (Brewer, Barton, & Dumoulin, 2014; de Haas, Schwarzkopf, Anderson, & Rees, 2014; Dumoulin, Hess, May, & Harvey, 2014; Dumoulin & Wandell, 2008; Haak, Cornelissen, & Morland, 2012; Harvey & Dumoulin, 2011; Papanikolaou, Keliris, Papageorgiou, Shao, & Krapp, 2014; Schwarzkopf, Anderson, Haas, White, & Rees, 2014; Verghese, Kolbe, Anderson, Egan, & Vidyasagar, 2014; Zuiderbaan, Harvey, & Dumoulin, 2012). The most commonly used stimuli in this approach – also referred to as population receptive field (pRF) mapping – are bar-aperture stimuli containing high-luminance contrast reversing patterns (the carrier) presented on a blank background. However, other apertures and carriers can be used for characterizing pRF properties as well (Alvarez et al. 2014, Dumoulin et al., 2014). One of the reasons for the popularity of high luminance contrast patterns is that they tend to drive a wide range of neurons – thus resulting in high signal-to-noise BOLD responses. However, this ability may also be a disadvantage for estimating detailed pRF information. Using non-selective stimuli will activate a large number of neurons in the population – with a wide range of receptive field properties – that will all contribute to the average response of a voxel. Using more selective stimuli will open the way for estimating more detailed pRF properties of a narrower population of neurons.

Here, we test this notion and turn to using local differences in orientation texture (to which we will refer to as orientation-contrast). The abundance of orientation selective neurons throughout early visual cortex is by now well established. However, although previous studies have shown that orientation processing can be studied using fMRI (Yacoub et al. 2008, Freeman, Brouwer, Heeger, & Merriam, 2011; Haynes & Rees, 2005; Kamitani & Tong, 2005; Swisher et al., 2010) the use of orientation information for characterizing the retinotopic specificity of BOLD signals is still uncommon.

We reasoned that we could use orientation-contrast as a carrier to define a bar-like aperture by placing similar small elements throughout the visual field albeit with elements having different orientations in the bar and the background (figure 1). As a consequence, both aperture and background will be continuously stimulated. However, the bar is visible only because of the orientation-contrast creating virtual edges. Therefore, we expect that our stimulus selectively targets only orientation-contrast selective neurons. To our knowledge, our study is the first to use a second-order orientation-contrast stimulus in combination with population receptive field (pRF) modeling to characterize the retinotopic specificity of BOLD signals in the human visual cortex.

# Methods

*Participants*

Prior to scanning, participants signed an informed consent form. The five participants (3 female, 2 male; average age: 24; age-range: 22-27) had normal vision. Our study was approved by the UMCG medical ethical review board.

*Stimulus presentation*

Visual stimuli were created using MATLAB and the Psychtoolbox (Brainard, 1997; Pelli, 1997). Stimuli were presented on an MR compatible display screen (BOLDscreen 24 LCD; Cambridge Research Systems, Cambridge, UK). The screen was located at the head-end of the MRI scanner. Participants viewed the screen through a tilted mirror attached to the 16-channel SENSE head mounted coil. Distance from the eyes to the screen (measured through the mirror) was 80 cm. Screen size was 36 x 23 degrees of visual angle.

*Stimulus Design*

*Luminance-contrast defined retinotopy (LCR)*

For the retinotopy scan, we presented a drifting bar aperture defined by high-contrast flickering texture (Dumoulin & Wandell, 2008; Harvey & Dumoulin, 2011; Zuiderbaan et al., 2012). The bar aperture moved in 8 different directions (four bar orientations: horizontal, vertical and the two diagonal orientations), with for each orientation two opposite drift directions). The bar consisted of alternating rows of high-contrast luminance checks drifting in opposite directions. The bar moved across the screen in 16 equally spaced steps each lasting 1 TR. The bar contrast, width and spatial frequency were respectively 100%, 2.5 degrees and 0.5 cycles per degree. After each pass and a half, 12 seconds of a blank stimulus at mean luminance was presented full screen.

*Orientation contrast defined retinotopy (OCR)*

The orientation-contrast defined retinotopy stimulus (OCR) was designed in such a way that the aperture size and shape and movement of the aperture approximately corresponded to that of the conventional LCR stimulus. By doing so, we preserved the spatial attributes of the visible larger-scale object (the bar aperture). However, instead of a blank grey background and a luminance contrast-defined bar, both the background and the aperture bar consisted of small oriented Gabor patches. Bar and background could be distinguished from each other on the basis of their different base orientations (Figure 1). The OCR stimulus consisted of a field of small Gabor patches (GB) that filled the entire screen. Gabors were positioned in a [90 x 56] grid covering the entire screen. The width and spatial frequency of the gabor patches were respectively 0.33 deg and 3 cycle per degree. Gabor center-to-center distance was 0.4 deg and sigma of the Gaussian envelope was 0.21 deg. Absolute orientation of the Gabors varied randomly and was refreshed every 125 ms. A relative difference in base-orientations between fore- and background of 45° revealed the bar aperture from the background. The bar moved across the screen in 20 equally spaced steps each lasting 1 TR. The bar width was identical to the LCR. The contrast of the individual Gabor elements was 30%. The position of each Gabor's center varied randomly between 0-0.06 deg to reduce adaptation. In the edge model (figure 1d) the width of the edge was 0.73 deg (the size of a Gabor plus the center-to-center distance). The code to generate OCR stimuli will be made available through www.visualneuroscience.nl

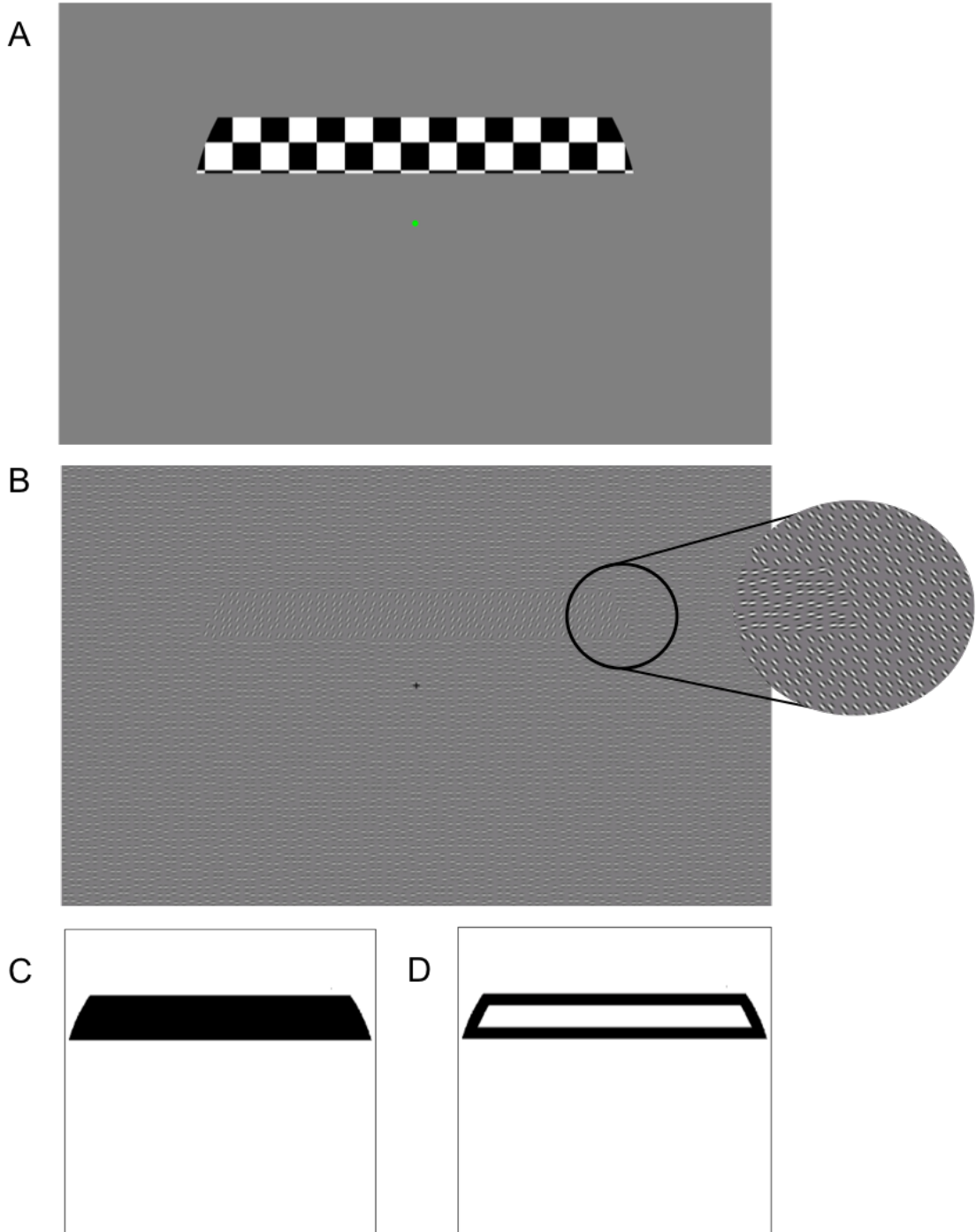

Figure 1: A: Example bar aperture stimuli for Luminance-Contrast (LCR) and B: Orientation-Contrast (OCR) defined Retinotopy. C: Stimulus model matrix used in the LCR and OCR-field

(OCRf) population receptive field (pRF) analyses. D: OCR-edge (OCRe) matrix used in an additional pRF analysis of the OCR data.

*MRI scanning*

*Scanner*

Scanning was carried out on a 3 Tesla Philips Intera MR-scanner using an 8-channel receiving SENSE head coil. A T1-weighted scan covering the whole-brain was recorded to chart each participant's cortical anatomy. The functional scans were collected using T2*-weighted echo-plannar imaging sequences, with a flip angle of 80°, a TR of 1.5 second and a TE of 30 ms, and a voxel size of 2.3 mm isotropic. Each functional scan consisted of 24 slices aligned parallel to the calcarine sulcus.

*Experimental procedure*

Participants were scanned using both the standard luminance-contrast defined retinotopy (LCR) and our new orientation contrast-defined retinotopy (OCR) in two different sessions of approximately 1 hour. For LCR (OCR) a single run consisted of 136 (188) functional images (duration of 204 s and 282 s respectively). Eight prescan images (duration of 12 s) were discarded.
In the first session, the anatomical scan and the LCR experiment (8 runs) were performed. In the second session, the OCR experiment (10 runs for four and 11 runs for one subject) was performed. During scanning, participants were required to perform a fixation task in which they had to press a button each time the fixation point turned from green to red. The average (std. err) performance on this task was 82% (±5%) for the LCR and 84% (± 4%) for the OCR.

*Preprocessing*

The functional imaging data were pre-processed and analyzed using the mrVista software package from Stanford University (http://white.stanford.edu). The T1-weighted whole-brain anatomical images were re-sampled to a 1 mm isotropic resolution. Automatic gray and white matter segmentation was carried out with FSL software (Smith et al., 2004) and subsequently edited manually. The cortical surface was reconstructed at the white/gray matter boundary and rendered as a smoothed 3D mesh (Wandell et al., 2007). The functional scans were within and between scans motion corrected and aligned to the first scan of every session. The anatomical and functional scans were coregistered.

*Population Receptive Field (pRF) modeling*

For both stimulus types (LCR and OCR), population receptive field analysis was performed on the functional MRI data (Dumoulin & Wandell, 2008). For each voxel, a 2D-gaussian model was fitted with parameters $x_0$, $y_0$, and $\sigma$ where $x_0$ and y0 are the receptive field center coordinates and σ is the spread (width) of the Gaussian signal, which is also the pRF size. We used SPM's canonical difference of gammas for the HRF model. All the parameter units are in degrees of visual angles and stimulus-referred.

We analyzed the responses to the OCR stimuli also using a second (edge) model, that assumes that only neurons in a region near the (virtual) border between the fore- and background of Gabor patches are activated by the stimulus (figure 1D).

*Statistical analysis*

Data was thresholded by retaining the pRF models that explained at least 20% (10% for the maps shown in figure 3) of the variance in the BOLD response and that had an eccentricity in the LCR analysis in the range of 2-9 degrees. Unless mentioned otherwise, for the analyses, results (i.e. pRF model parameters) were binned over eccentricity, in 1-degree bins, separately for each hemisphere. Statistical analysis was performed using repeated measures ANOVA, with ROI, pRF model (LCR, OCR), hemisphere (LH, RH) and eccentricity-bin as within subject parameters. Due to the modest study population size, participant was not treated as random effect. A p-value of 0.05 or less was taken to indicate significant results.

# Results

In this study, we use second order orientation-contrast defined stimuli to perform retinotopic mapping (Orientation Contrast Retinotopy, abbreviated as OCR) and use population receptive field (pRF) modeling to analyze the data. We compare results to Luminance Contrast Retinotopy (LCR). To preview our main results, we find that in general, pRF sizes obtained with OCR are smaller than those obtained with LCR. This is most notable in ventral cortical areas (LO1 and LO2). Moreover, while there are small shifts in the pRF locations in early visual cortex (V1-V4), in the higher order ventral areas (LO1 and LO2) we find substantially lower pRF eccentricities. EV as a function of eccentricity tended to be lower but more stable for OCR than for LCR. OCR and LCR derived maps of visual cortex are highly comparable.

*Comparison of BOLD signals evoked by LCR and OCR stimuli.*

Figure 2b shows representative time-series for OCR and LCR stimuli. Figure 2c shows the average standard deviation of the signal in various ROIs. It is clear that the signal modulation evoked by OCR stimuli is smaller than that evoked by LCR stimuli. Remarkably, for LCR, the signal variation (standard deviation) is markedly smaller for higher (LO1, LO2) than for lower order (V1-4) areas. The low response modulation to the LCR type stimulus in these areas is a common finding (Dumoulin and Wandell, 2008, Brewer et al., 2005; Larsson and Heeger, 2006; Wandell et al., 2005). For OCR, signal variation is comparatively lower but stable over areas.

*Comparison of LCR and OCR derived retinotopic maps*

Figure 3 shows polar angle, eccentricity and pRF size maps obtained for LCR and OCR stimuli projected onto the inflated hemisphere of a representative participant. This visualization shows that overall, the maps and ROI borders obtained with OCR and LCR are very comparable. For OCR, it can be appreciated that in the extrastriate regions, pRF sizes are generally smaller (less red).

a.

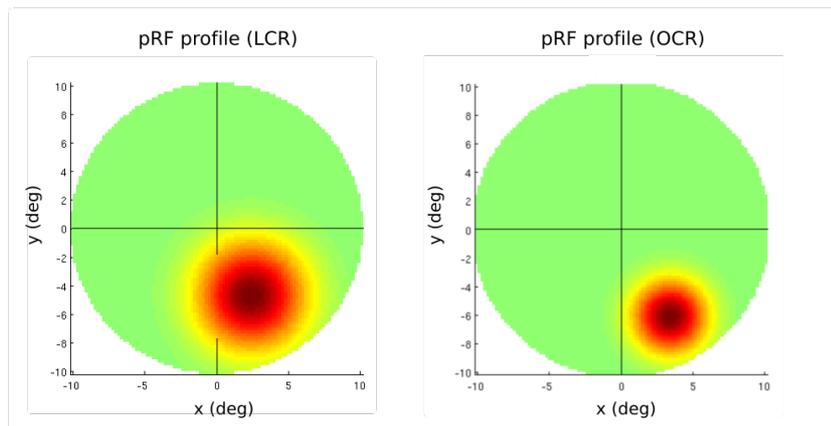

b.

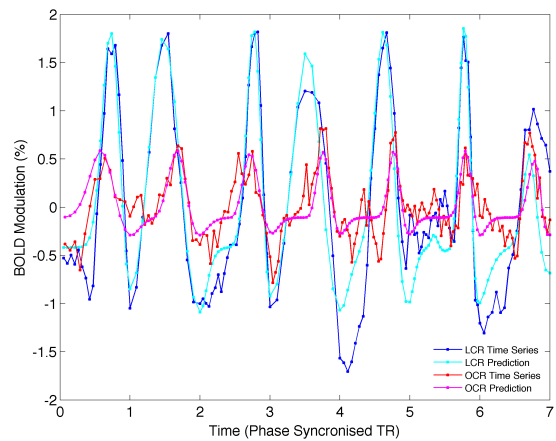

c.

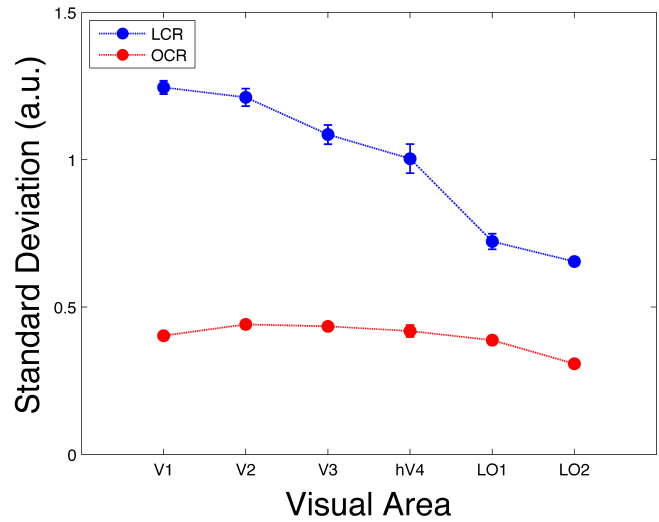

Figure 2: a) Coverage maps showing the location and size of the pRF for a single representative V1 voxel as determined for the LCR (left) and OCR (right) condition. The coverage map shows the region in visual space to which the voxel responded, as determined by pRF modeling. b): BOLD time-series and pRF model predictions for the same V1 voxel during the presentation of LCR (data: blue; fit: cyan) and OCR (data: red; fit: pink) stimuli, c): Mean of the standard deviation of the BOLD time series (in arbitrary units) for different ROIs for LCR (blue) and OCR (red) stimuli. Average results for 5 participants, 2 hemispheres each. Error bars indicate the standard error of the mean over hemispheres.

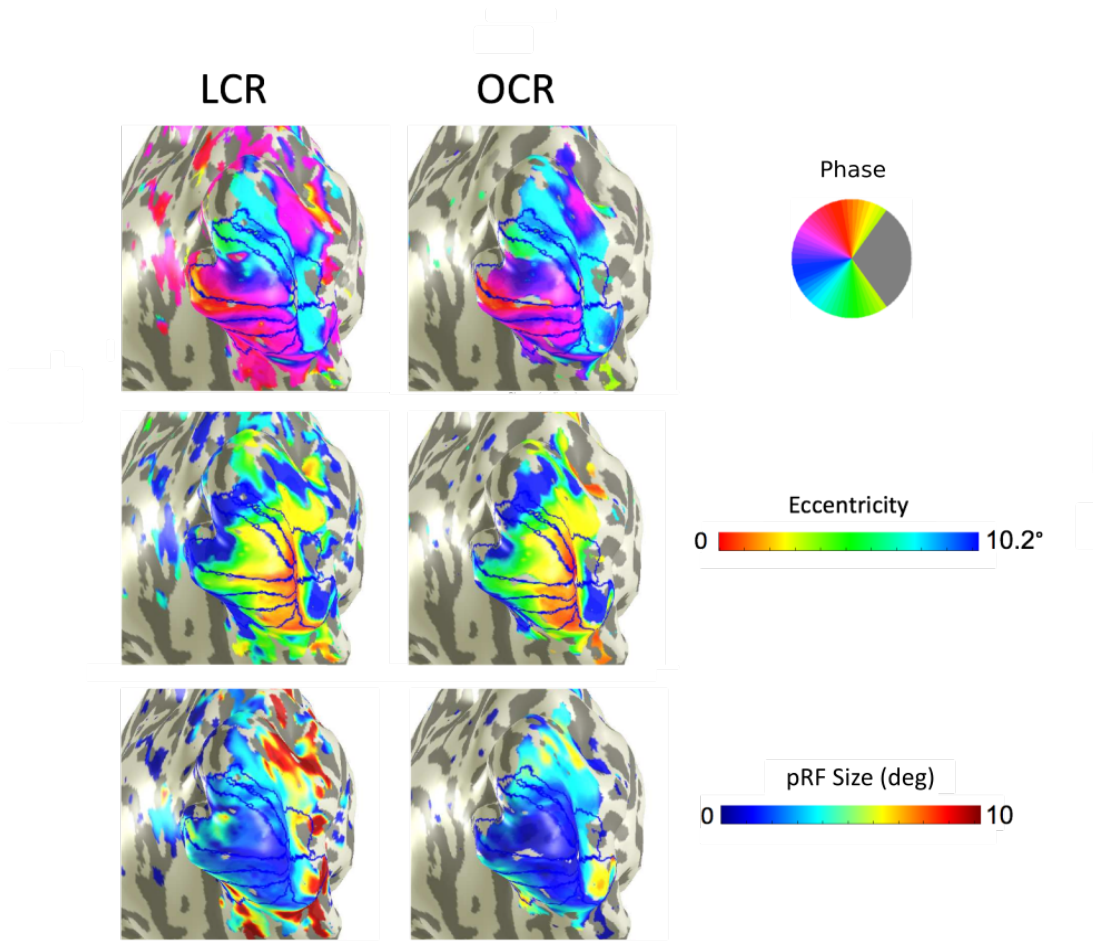

Figure 3: Polar angle, eccentricity and pRF size maps for the right hemisphere of participant S04 obtained with luminance- and orientation-contrast (field) defined stimuli. In all cases, the explained variance threshold was set to 0.1. Dark blue lines indicate ROI borders. The maps for the field and edge model analysis of the orientation-contrast defined stimuli were nearly identical and are therefore not shown separately.

*Comparison of voxel-wise pRF eccentricities for LCR and OCR*

Figure 4 compares the eccentricities assigned to each voxel based on the results of the pRF analysis in the OCR and LCR model conditions. Most notable in foveal V1 and V2, the eccentricities assigned by OCR are somewhat higher than those for LCR. In V3 and V4, the eccentricities assigned are quite comparable, evident from the relatively small deviation from oblique. For LO1 and LO2, the eccentricities assigned differ substantially, with assigned eccentricities generally being smaller for OCR. There was a significant three-way interaction between ROI, condition and eccentricity ($F(35, 140)=3.99$, $p<0.01$).

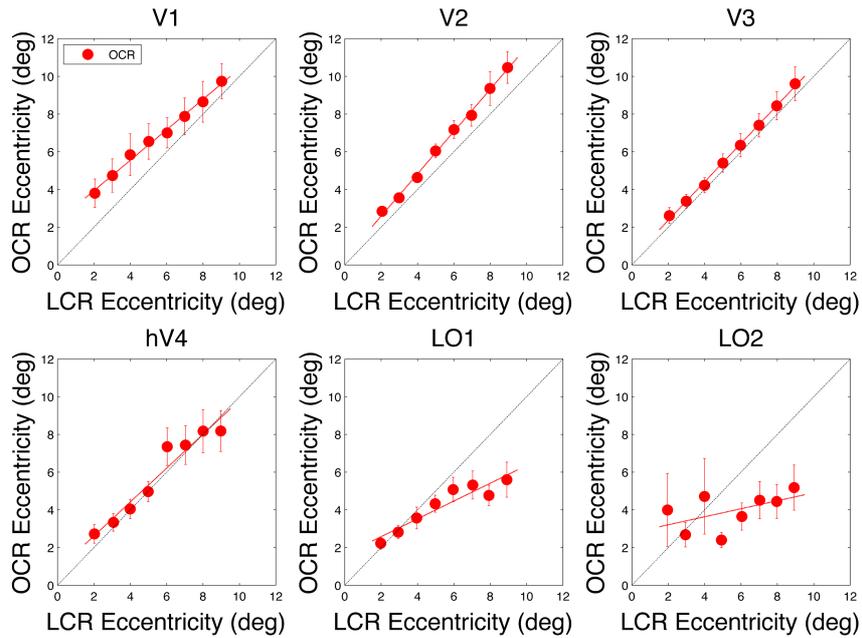

Figure 4: Eccentricity assigned to each voxel based on OCR and LCR for 6 different ROIs. LCR-assigned pRF eccentricity is shown on the x-axis. while eccentricity assigned based on the OCR model analysis is shown on the y-axis. Average results for 5 participants, 2 hemispheres each. Error bars indicate the standard error of the mean over hemispheres.

*Comparison of voxel-wise pRF sizes for LCR and OCR*

Figure 5 shows pRF sizes estimated based on the pRF modeling in the LCR and OCR condition, as a function of the eccentricity assigned by each of the respective model analyses. As is commonly observed, pRF size increases with eccentricity. This was comparable for all conditions and ROIs. In general, the pRF sizes determined based on the OCR analyses were somewhat smaller than those determined based on the LCR analysis. The difference in pRF size between conditions was most prominent for V4, LO1 and LO2. The three-way interactions between condition, ROI and eccentricity ($F(35, 140)=4.23$, $p<0.01$) was significant, indicating that the difference in estimated size is larger for higher order areas.

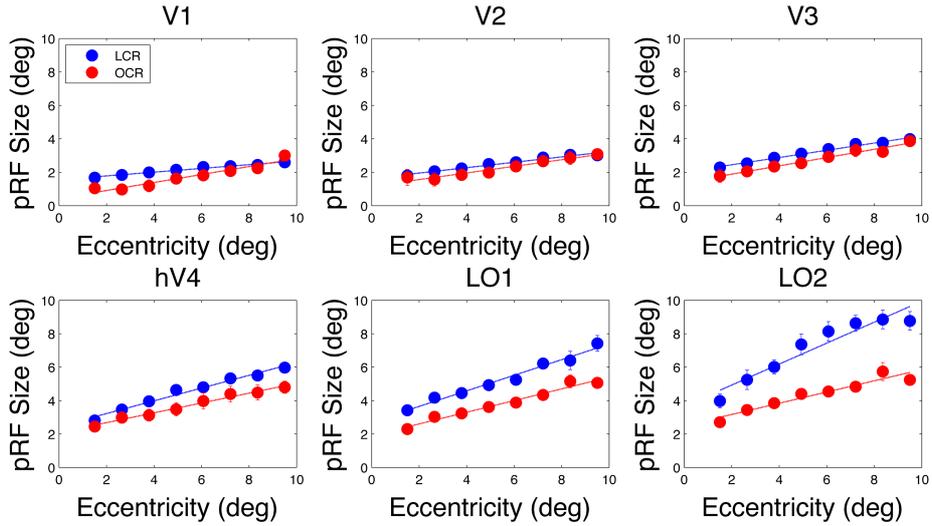

Figure 5: Average pRF size plotted as a function of pRF eccentricity (as assigned by each respective model analysis) for six different ROIs. Eccentricity was binned in bins of 1 deg. Average results for 5 participants, 2 hemispheres each. Error bars indicate the standard error of the mean over hemispheres.

*Comparison of Explained Variance for LCR and OCR-based analyses*

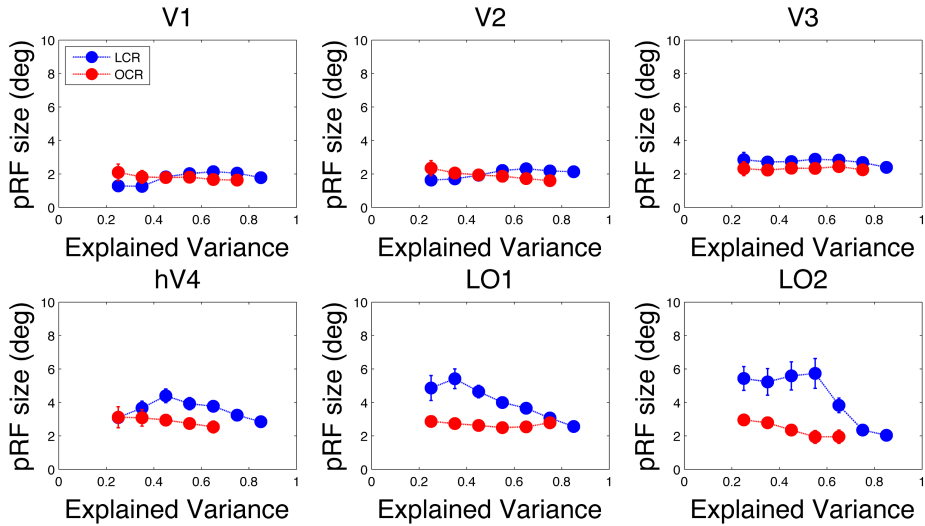

Figure 6: (a) pRF size as a function of explained variance (EV) for six different ROIs. EV was binned in bins of size 0.1. Each bin shows the average results for 5 participants, 2 hemispheres each. Note: OCR misses data points in some ROIs, as there were no models with this level of EV. Error bars indicate the standard error of the mean over hemispheres.

Figure 6 shows pRF size as a function of EV. For the early visual areas, pRF size does not appear to depend on EV. However, for the extrastriate areas hV4, LO1 and LO2, there is a difference between LCR and OCR. For LCR, pRF size does depend on EV, with lower EV models resulting in larger pRFs in particular for high order areas. This dependence of pRF size on EV is not observed for the OCR stimulus.

Figure 7 summarizes several of our results over ROIs. For the OCR data, the results for an additional analysis using a different stimulus model (OCR edge) were added. The pRF size determined for OCR and LCR differed for the various ROIs (F(5, 20)=16.2 p<0.01). While pRF size increases monotonically from lower to higher order visual areas for both LCR and OCR, this effect was most pronounced for LCR (see figure 7A).

The edge model estimated the smallest pRF sizes, on average. Figure 7B plots the same data, but in terms of an average reduction in pRF size. It shows that the use of OCR resulted in average pRF size reductions of up to 38%. Finally, figure 7C shows the average EV in each ROI. The average EV determined for the different conditions differed for the various ROIs. Using different models for the OCR data did not affect EV. For LCR, EV decreased for higher order areas compared to V1-V3. For OCR, EV increased slightly from V1 to V3, and peaked in LO1.

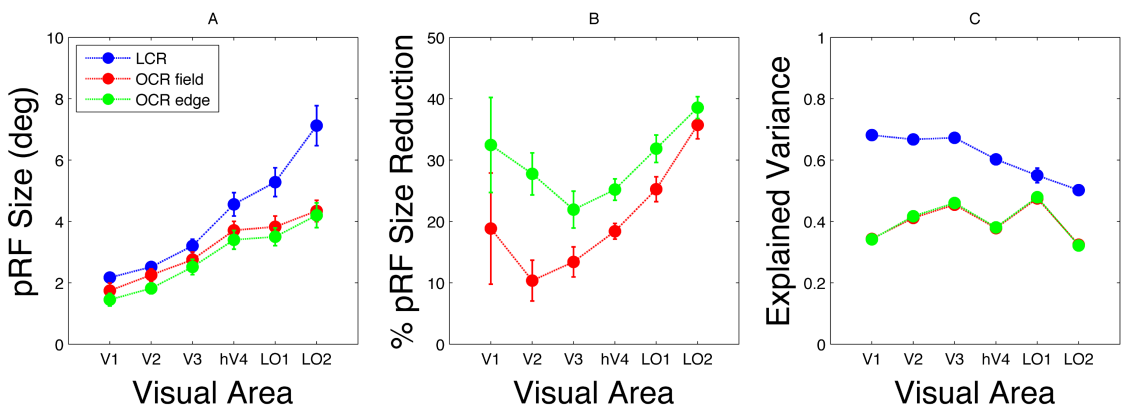

Figure 7: Summary of average results for the different ROIs. Also the results of an additional analysis for the OCR data using a different model (OCR edge, figure 1D) are shown. A: Comparison of pRF sizes. For each observer and ROI, pRF size was averaged over the eccentricity bins as shown in figure 5. B: Same data but plotted in terms of pRF size reduction ((LCR-OCR)/LCR*100%). C: Average explained variance for each ROI. For each ROI, explained variance was averaged over eccentricity bins. Error bars show standard error of the mean over hemisphere.

## Conclusion

Our results thus far suggest that using the OCR stimulus results in pRF estimates that are distinctive from those obtained using the standard LCR, in particular for higher order areas LO1 and LO2. However, before being able to attribute this to more selective stimulation, a number of control conditions are required. The OCR stimulus differed from the LCR stimulus not only in the application of orientation contrast, but also in a number of low-level stimulus aspects. Therefore, we performed a control experiment aimed at verifying the possible influence of the-se low-level stimulus factors on the estimation of pRF properties. We examine the potential influence of: 1) a reduction in contrast energy of the stimulus, 2) the presence of a continuous background, and 3) an increase in the stimulus' main spatial frequency.

## Methods (Control Experiment)

Unless indicated otherwise, the methods were similar to those described previously.

*Participants*
The control experiments were performed by 2 of the original participants that were also included in the main experiment (1 female, 1 male, ages 24 and 26).

*Stimulus Design*
We created three variants of the LCR bar stimulus (depicted in figure 8). a. The standard LCR stimulus, but rendered at a much lower luminance contrast (2%).
b. A standard high-contrast LCR with smaller checks with a main spatial frequency more comparable to that of the OCR stimulus (4 cycles per deg).
c. The standard LCR stimulus to which a continuous dynamic high-contrast (100%) white noise background.

*Experimental procedure*
The participants were scanned using the standard and the various new LCR stimuli in 2 different sessions of approximately 1.5 hours each. In each session, the different stimuli were presented in a pseudo-random order.

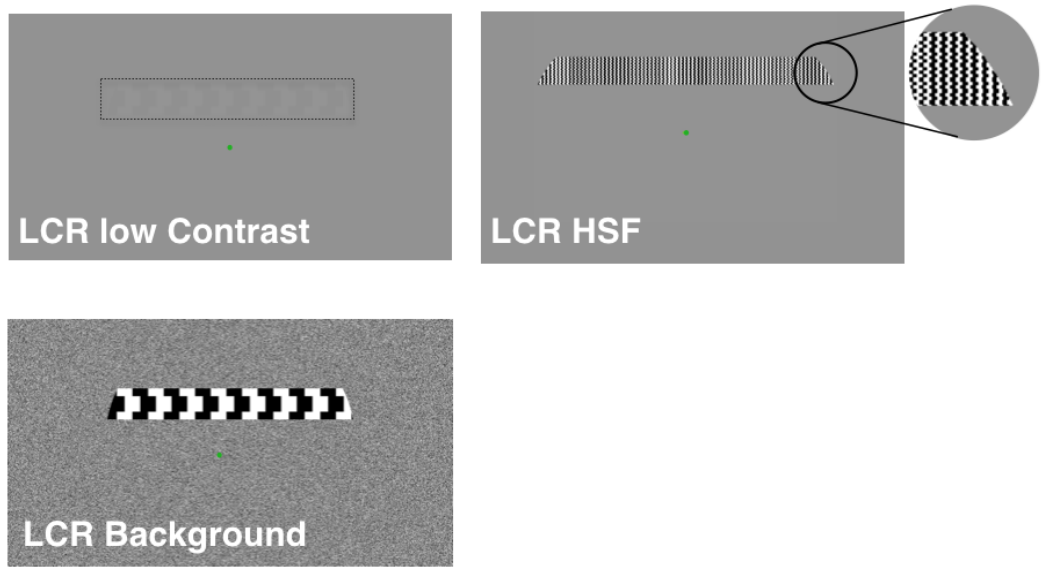

Figure 8. Additional stimuli used in the control experiment. A: Low contrast LCR (2% contrast) B: High spatial frequency LCR (100% contrast, 4 cycles per degree). C: LCR with continuous dynamic high contrast (100%) white noise background present.

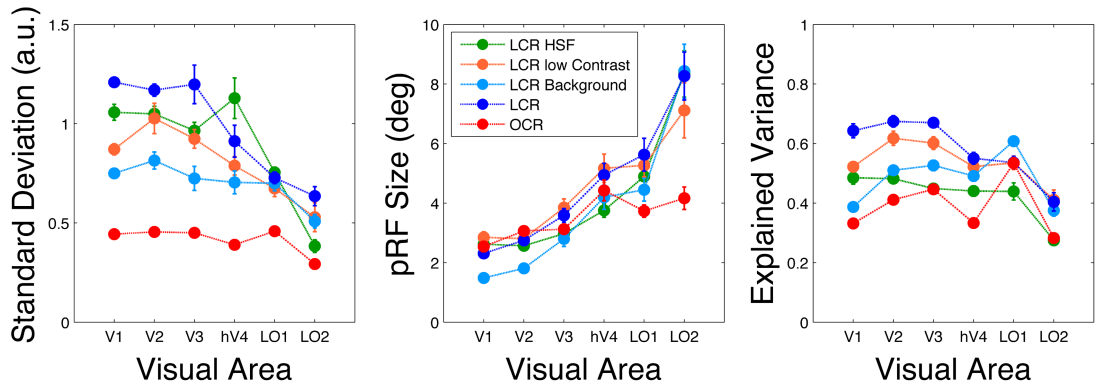

Figure 9: Summary of results for the control experiment using 5 different stimulus paradigms. A: Mean of the standard deviation of the BOLD time series (in arbitrary units) for different ROIs. B: Comparison of average pRF size in the different ROIs. C: Comparison of average EV in the different ROIs. Average results for 2 participants, 2 hemispheres each. Error bars indicate the standard error of the mean over hemispheres.

### Results and Conclusion (Control Experiment)

Figure 9 summarizes the results for the control experiment. The average standard deviation of the BOLD signal for the various LCR paradigms varied substantially. Despite this, the average estimated pRF sizes did not differ substantial for most of the LCR stimulus variants. However, as in the main experiment, in LO1 – and LO2 in particular – the average pRF size was substantially lower for the OCR than for most LCR variants. In V1- LO1, but not LO2, adding a noise background to the LCR stimulus also resulted in smaller pRF sizes. The difference in EV between stimulus paradigms was substantial. Again, we also see that EV is more stable over ROIs for OCR than for most LCR variants. Importantly, however, there is no obvious relationship between the average signal standard deviation, the model EV and the pRF size estimated for the various ROIs. We conclude that OCR results in smaller pRF sizes in area LO1 and LO2, which cannot be accounted for by low-level stimulus factors (such as spatial frequency, contrast and presence of background) or signal-to-noise of the BOLD signal.

### Discussion

Visual field maps derived from OCR and LCR were qualitatively similar. Only small and non-systematic differences in border locations were present. Hence, for determining the general layout of visual cortical maps, OCR holds no clear advantage over LCR. However, in general, the estimated pRF sizes were meaningfully smaller for OCR compared to LCR. In addition, for LO1 and LO2, the voxel-wise estimated pRF eccentricities were generally lower for OCR than for LCR. Below, we will discuss the origin of these differences in pRF estimates and argue they are desirable – despite the lower EV for OCR. Therefore, we conclude that using our approach, it is possible to selectively target particular neuronal populations, opening the way to use pRF modeling to dissect the response properties of more clearly-defined neuronal populations in different visual areas. Below, we discuss this in more detail, as well as the implications for understanding pRF modeling and cortical organization and function.

*OCR selectively targets orientation-contrast sensitive neurons*

The pRF modeling based on OCR generally assigned smaller pRF sizes to voxels. One reason for this might be that the estimated pRF size depends on the signal modulation (signal-to-noise ratio) of the BOLD signal. Can differences in signal modulation also explain differences in estimated pRF size?
First, in our control experiment a number of LCR variants were used that also resulted in a substantially reduced BOLD signal modulation (figure 9a). However, overall, this did not obviously affect pRF size estimation (figure 9b). So, this general observation argues against a direct link. The supplementary information (figure S1) contains a further analysis that also argues against this link.

To further answer this, we will assess the possible causes of the reduced signal modulation. Unlike in LCR, in OCR the entire stimulus was continuously stimulated in an identical manner for bar and background – except for Gabor orientation. Three mechanisms might explain the reduced BOLD signal modulation in OCR, as we observed. The first mechanism is a hemodynamic one. The stimulation will evoke a BOLD response in a large cortical region (i.e. the representation of the entire stimulus) thereby limiting the amount of blood available to the cortical region stimulated by the bar aperture only. Therefore, background stimulation will lead to a decrease in BOLD amplitude. This process is known as "hemodynamic stealing" (Shulman et al, 1997; Allison et al, 2000; Shmuel et al 2002; Olman et al., 2007).

In the control experiments, the addition of a background to the LCR stimulus had a substantial effect on signal modulation in areas V1-V4, but little to no influence in LO1 and LO2 (figure 9a). In areas V1-L01, the reduced signal modulation is accompanied by a reduction in estimated pRF size (figure 9b). However, in particular for LO2, despite a comparable signal modulation for the OCR and LCR-with-background conditions, the difference in estimated pRF size is very large. This suggests that the reduced signal modulation by itself is not the cause of the smaller estimated pRF size.

The second mechanism is a neural one, and is generally referred to as surround suppression (also referred to as lateral inhibition). Surround suppression influences the activity of a neuron due to the activity of neighboring neurons and also affects the magnitude of the BOLD response (Zenger-Landolt and Heeger, 2003, Chen 2014, Williams et al, 2003 and Pihlaja et al, 2008). Depending on the similarity of stimulus and background (e.g. in terms of orientation), surround suppression can either enhance or reduce responses. For OCR, given the difference in orientation between the bar and the background, it would be expected to increase the amplitude and sharpness of the neural response to the texture boundaries. In accordance, the edge model did indeed result in slightly smaller pRFs than the field model. However, if surround suppression were an important cause of the reduced pRF size observed for OCR, we would have expected that the edge model would also have provided a significantly better fit to the data. This is not the case (figure 7c). This suggests that surround suppression is not a likely explanation for the observed effect.

A third and final mechanism is based on the population characteristics of the pRF and the selectivity of the OCR stimulus. The pRF measures the collective RF of all active neurons in a voxel. Our second-order stimulus only drives a subset of those neurons – namely the ones that are sensitive to orientation contrast. First, driving a subset of neurons will result in a lower signal modulation. This is expected, given the correlation between the size of the estimated receptive field and the volume of integrated activity (Land et al. 2013). Second, it is quite conceivable that this subpopulation has a smaller and less scattered collective RF than that of all neurons that are driven by the LCR stimulus – which would include orientation-selective and non-selective neurons, luminance- and luminance-contrast sensitive neurons as well as e.g. motion sensitive neurons. Given that the other explanations fall short, we deem this latter explanation the most likely one.

*Desirable features of OCR*

A feature of OCR was that the signal modulation and the EV of the pRF models – while lower than for LCR – remained relatively stable over ROIs. In contrast, the LCR pRF models tended to show a decrease in EV when moving up the visual hierarchy. Hence, although the EV of OCR models are lower – amongst others related to the lower signal modulation – the models tended to have very comparable explanatory value for all cortical ROIs. We consider this an advantage when comparing functional properties over different cortical areas.

For early visual areas, pRF size did not depend on model EV for either OCR or LCR. However, for LCR in higher order areas, we found that pRF size does depend on EV, with models with a lower EV resulting in substantially larger pRFs (figure 6). Given that on average EV tended to be lower for higher order areas (figure 7c), potentially, this could result in overestimating pRF sizes in these areas when using LCR. For OCR – also in the LOs – pRF size was largely independent of EV, which we consider another desirable feature of this paradigm.

The work of Olman, Inati & Heeger, 2007 suggests why using a second order stimulus may result in spatially more accurate pRFs. They found that large pial veins can result in a displacement of the BOLD signal in the stimulated cortical region. They also found that displacement was reduced when the stimulus was alternated with its complement. During pRF mapping BOLD displacement may lead to mislocalization and incorrect estimation of pRF properties. In a similar fashion, continuous stimulation of both the bar and the background – which can be regarded as the simultaneous presentation of the bar stimulus and its complement– may also reduce this BOLD displacement and improve the retinotopic spatial accuracy. Reduced BOLD displacement may at least partly explain why also the LCR with a continuous background resulted in a reduction in pRF size in various areas (albeit not LO2). Therefore, part of the effect of using a second order stimulus may be due to the addition of a continuous background, yet it can-not explain the full range of our present findings in higher order areas. Irrespective, even though the exact mechanisms remains a bit elusive, it means that using a second order type of stimulus will improve accuracy of the pRF estimates.

Estimating smaller pRF sizes may also improve spatial accuracy. As we argue in the supplementary information (figure S2), it may reduce the influence of cortical magnification on pRF mapping. We conclude that – in particular for studying the pRF properties of higher order visual areas and for inter-area comparison of pRF properties – the second order OCR paradigm may hold advantages over the standard LCR paradigm. However, we also note that characterization of high order visual areas would require mapping many more dimensions than just orientation contrast tuning.

*Relevance for understanding cortical function*

Conventional retinotopic mapping and pRF modeling in the current form have been developed and tested for luminance based stimuli. Thus these methods measure a luminance-contrast related spatial organization in the brain. Although the luminance sensitive neurons dominantly exist in the visual cortex of primates (Kimoshita and Komatsu, 2001), these hardly represent the entire neuronal 'population' in the visual cortex. For this reason, one has to be careful when mapping the visual field locations and take into account what type of stimulus these maps are being compared to. Furthermore, estimating the pRF properties for a selective population of neurons in different cortical areas might have implications for understanding the function of these areas. For example, based on the OCR data, the orientation selective voxels in LO ROIs appear to process much smaller sections of the visual field (lower eccentricity estimates), and with much higher resolution (smaller pRF sizes). In other words, whereas on the basis of predominantly luminance-contrast driven voxel activity these ROIs would be described as areas that coarsely analyze a fairly substantial section of the visual field, the OCR based analysis suggests that these ROIs actually scrutinize the fovea and parafovea. This makes sense in the context of their presumed role in object recognition (Grill-spector, Kourtzi, & Kanwisher, 2001). Still, we should point out that pRF size and eccentricity has only been revealed for the orientation contrast sensitive sub-population of neurons, with no indication that this represents the spatial tuning of a majority of the neurons.

A different way to probe cortical function is by using different models to analyze the same functional data. OCR stimuli may potentially inform about the early stages of object-related processing. For this goal, we tested two models. In the OCR edge analysis, the assumption was that the primary signals are evoked by the local differences in orientation at the edges. In the OCR field analysis, the assumption was that fore- and background signals are based on stronger grouping for similarly oriented Gabors (e.g. Parkes, Lund, Angelucci, Solomon, & Morgan, 2001). For this reason, we also included the aperture's surface in the definition of the stimulus. EV was highly similar between the two OCR models as were the maps obtained with the two OCR models. pRf eccentricity estimates were highly comparable between models, while only pRF size estimates were somewhat smaller for the edge model, in particular in areas V1-V3. Consequently, although results were slightly different, neither model provided a superior explanation for the signals evoked by the orientation-contrast stimuli. Perhaps using stimuli with wider bars may enable a better distinction between these models.

The EV for the LCR-based pRF models (as well as the standard deviation of the raw BOLD signal), dropped for V4 and the LOs compared to that for V1-V3. In contrast, for OCR, the EV remained relatively stable over areas and even peaked in LO1. This suggests a more equivalent sensitivity of early and later regions to second order stimuli, such as those based on orientation contrast. This finding corroborates a patient study that indicated that V1 is sufficient to process simple orientation discrimination tasks but that ventral extra-striate regions are required to properly detect texture boundaries (Allen, Humphreys, Colin, & Neumann, 2009).

*Limitations*

The Gabor patches in the current experiment were all of similar size. In a recent study, eccentricity scaling had a significant effect on goodness of fit and pRF size estimates (Alvarez, de Haas, Clark, Rees, & Schwarzkopf, 2015). Future experiments could consider scaling the Gabor patches with eccentricity. It also remains to be determined conclusively whether the eccentricity differences for OCR and LCR are a consequence of signal leakage. More accurate pRF models that take BOLD leakage into account could also prove useful in this realm.

For logistic reasons, the LCR and OCR paradigms presently used were very similar, but not identical in terms of scan and run duration. Future experiments could consider using identical spatio-temporal profiles for both stimulus types. The signals evoked by LCR and OCR differed in BOLD amplitude. The lower amplitude signal for OCR was somewhat offset by a longer overall sampling duration for OCR than for LCR. Nevertheless, even though there were a few more and somewhat longer OCR sessions, this still resulted in models with lower explained variance. Future experiments could consider using lower contrast LCR stimuli to equalize this aspect and evaluate the consequences thereof on EV and pRF property estimates.

*Future studies*

Our current pRF analyses were based on the assumption that the orientation contrast of the Gabors in the OCR stimulus forms illusory edges and that the Gabors are grouped into objects based on the orientation differences. More detailed – and therefore potentially more informative – models could be created that take the actual orientation of each Gabor in each frame into account when modeling the BOLD responses. Such detailed models could also be used to explicitly predict both excitatory and inhibitory responses e.g. based on saliency models (Zhaoping 2008). Moreover, also periods of blank backgrounds (i.e. Gabors only present in or outside the bar aperture or no Gabors present at all) could be in-cluded. This could result in a stimulus that integrates the properties of the current OCR and LCR stimuli.

Our finding that adding a dynamic background to a conventional retinotopic mapping stimulus does reduce pRF size in early visual areas is an interesting secondary finding that warrants further study. As this influence was completely absent in LO2, the reduction appears to be caused for different reasons than the reduction in size observed for the OCR. Irrespective, researchers interested in improving the resolution of pRF estimation in early areas – yet preferring to continue using a more conventional stimulus – could consider making this relatively straightforward change to their paradigm.

The here presented second-order stimulus is very versatile. Other visual aspects could be tested for by using Gabor patches at various levels of orientation-contrast, various spatial or temporal frequencies, chromatic differences or other aspects that could result in figure-ground segmentation. By varying Gabor similarity and inter-Gabor distance, the Gabor fields could also be used to study feature integration and crowding. Finally, at higher fMRI field strengths, the laminar layers that evoke OCR related BOLD responses could be studied, and whether this differs from the classical LCR (De Martino et al., 2013; Muckli et al., 2015).

*Conclusion*

Population receptive field properties estimated based on a second-order orientation-contrast defined stimulus differ from those obtained using a classic luminance contrast defined stimuli. Estimation of neuronal properties is crucial for interpreting cortical function. Therefore, we conclude that our approach – despite evoking somewhat lower amplitude response – can be used to selectively target particular neuronal populations, opening the way to use pRF modeling to unravel the response properties of clearly-defined neuronal populations in different visual areas. Note that we are not advocating that our new stimulus should replace the traditional high-contrast stimulus in all cases. In our view, our work points out that one should carefully consider what aspect(s) of visual processing one wants to study and use (or design) a mapping stimulus that best characterizes that aspect. This may be the traditional luminance-contrast stimulus, our new orientation-contrast stimulus, or a different stimulus altogether. Finally, our control experiments demonstrate that the standard retinotopy paradigm appears rather robust to changes in contrast and spatial frequency of the stimulus. Depending on the goal of one's study, this can be considered an advantage or a limitation.

## Acknowledgments

FY was supported by The Graduate School of Medical Sciences (GSMS), University of Groningen, The Netherlands. JC was supported by the European Union's Horizon 2020 research and innovation programme under the Marie Sklodowska-Curie grant agreement No 641805. FWC was supported by the Netherlands Organization for Scientific Research (NWO Brain and Cognition grant 433-09-233).